\def \hg {{\hat g}}

\newcommand{\rf}[1]{(\ref{#1})}
\newcommand{\beq}{\begin{equation}}
\newcommand{\eeq}{\end{equation}}
\newcommand{\bea}{\begin{eqnarray}}
\newcommand{\eea}{\end{eqnarray}}
\newcommand{\beas}{\begin{eqnarray*}}
\newcommand{\eeas}{\end{eqnarray*}}

\documentstyle[12pt,epsf]{article}
\baselineskip=\normalbaselineskip
\ifx\TwoupWrites\UnDeFiNeD\else\target{\magstepminus1}{11.3in}{8.27in}
	\source{\magstep0}{7.5in}{11.69in}\fi
\newfont{\fourteencp}{cmcsc10 scaled\magstep2}
\newfont{\titlefont}{cmbx10 scaled\magstep3}
\newfont{\authorfont}{cmcsc10 scaled\magstep1}
\newfont{\fourteenmib}{cmmib10 scaled\magstep2}
	\skewchar\fourteenmib='177
\newfont{\elevenmib}{cmmib10 scaled\magstephalf}
	\skewchar\elevenmib='177
\makeatletter
\newcommand\nonsequentialeqnum{
	\@addtoreset{equation}{section}
	\def\theequation{\arabic{section}.\arabic{equation}}}
\newif\ifp@bblock  \p@bblocktrue
\newcommand\nopubblock{\p@bblockfalse}
\newcommand\topspace{\hrule height 0pt depth 0pt \vskip}
\newcommand\p@bblock{\begingroup \tabskip=\hsize minus \hsize
	\baselineskip=1.5\ht\strutbox \topspace-2\baselineskip
	\halign to\hsize{\strut ##\hfil\tabskip=0pt\crcr
	\the\Pubnum\crcr\the\date\crcr}\endgroup}
\renewcommand\titlepage{\ifx\TwoupWrites\UnDeFiNeD\null\vspace{-1.7cm}\fi
\vskip0.6cm
	\ifp@bblock\p@bblock \else\hrule height 0pt \relax \fi}
\makeatother
\newtoks\date
\newtoks\Pubnum
\newtoks\pubnum
\Pubnum={
FIT-HE-96-82 \crcr
hep-th/xxxxxxxx
}
\date={\today}
\newcommand{\frontpageskip}{\vspace{12pt plus .5fil minus 2pt}}
\renewcommand{\title}[1]{\frontpageskip
	\begin{center}{\titlefont #1}\end{center}\par}
\renewcommand{\author}[1]{\frontpageskip\par\begin{center}
	{\authorfont #1}\end{center}
	\nobreak
	}

\newcommand{\address}[1]{\par\begin{center}{\sl #1}\end{center}\par}

\renewcommand{\thanks}[1]{\footnote{#1}}
\renewcommand{\abstract}{\par\frontpageskip\centerline{\fourteencp Abstract}
	\vspace{8pt plus 3pt minus 3pt}}
\begin{document}
\pubnum{93-37}
\date{December 1996 \crcr}
\titlepage

\renewcommand{\thefootnote}{\fnsymbol{footnote}}
\title{
Field condensation and non-critical string for c$>$1\\
}

\author{
K.\ Ghoroku \thanks{
e-mail address: gouroku@dontaku.fit.ac.jp}
}

\address{

Fukuoka Institute of Technology\\
Wajiro, Higashi-ku, Fukuoka 811-02, Japan \\
}

\renewcommand{\thefootnote}{\arabic{footnote}}
\setcounter{footnote}{0}
\newcommand{\cleqn}{\setcounter{equation}{0} \indent}
\renewcommand{\theequation}{\thesection.\arabic{equation}}
\newcommand{\beqa}{\begin{eqnarray}}
\newcommand{\eeqa}{\end{eqnarray}}
\newcommand{\eq}[1]{(\ref{#1})}

\begin{abstract}
  Quantum theory of 2d gravity for $c>1$ is examined as a non-critical
string theory by taking account of the loop-correction of 
open strings 
whose end points are on the 2d world surface of the closed string. 
This loop-correction
leads to a conformal anomaly, and we obtain a modified target-space action
which implies a new phase of the non-critical closed-string. In this phase,
the dual field of the gauge field, which lives on the boundary, condenses
and the theory can be extended to $c>1$ without any instability.

\end{abstract}

\newpage

\renewcommand{\thesection}{\arabic{section}.} 
\renewcommand{\theequation}{\thesection \arabic{equation}}
\newcommand{\scs}{\setcounter{equation}{0} \setcounter{section}}
\def\req#1{(\ref{#1})}
\setcounter{footnote}{0}

\section*{1. Introduction}\scs{1}

The quantized 2d gravity can be formulated as
a conformal invariant nonlinear sigma-model
on the 2d manifold \cite{ddk}.
The non-critical string theory is also formulated
in the same way since it can be regarded as a
2d gravity coupled to several scalar-fields with conformal coupling, where 
the number ($c$) of the scalar-fields is related to the dimension of 
the target-space as $d=c+1$. 

The vacuum of the theories is determined by solving the conditions
of zero $\beta$-functions of the nonlinear sigma-model by the 
$\alpha'$-expansion \cite{clty}. In the sence of $\alpha'$-expansion,
some exact solutions
are given according to a simple ansatz \cite{wzw}
other than the well-known linear dilaton vacuum.
Quantum fluctuations of matter fields on these vacua have been
examined, and we found that
the properties like the renormalization group equations of matter-sector
are insensitive to the details of the vacuum configurations \cite{poly}.
This result 
could be understood such that the properties like the renormalization
group equations are determined by the short distance behaviors
on the 2d manifold and they do not depend on the global aspect of the vacuum.

While
there is a serious problem called as $c=1$ wall which means that the 2d
manifold becomes unstable for $c>1$, where 
$c$ is the central charge of the theory. 
This instability is observed as the complex string susceptibility
of the surface, and the numerical simulations \cite{ambj} indicate
the branched polymer phase of 2d manifold for $c>1$. 
This surface instability would heavily depend on the structure of the vacuum.
In terms of the non-critical string theory, the ground state
of the string-state becomes tachyonic, then the theory is unstable
for $c>1$ or $d>2$. And this difficulty could
not be removed in any vacuum state found until now.
In other words, any consistent theory of 2d gravity is not found still
for $c>1$ except for the special cases, $c=$7, 13, and 19 \cite{gerv}.
But the cases of $c=$7,13 and 19 do not include the physical degrees of
freedom and they are topological theory in this sence.
So, to try resolving this problem is very challenging.

It is possible to obtain a real string susceptibility for $c>1$ by adding
the curvature-squared term \cite{naka} to the world-sheet action or 
by considering a scalar-field which couples non-minimally to the 
scalar-curvature of the world sheet \cite{cham}. The reason 
why the complex susceptibility is avoided is that
the high-curvature configuration is suppressed due to these
terms, but these theories sacrifice the unitarity. On the one hand, the
touching interactions have been examined \cite{touch} in the recent matrix
model. This interaction can be regarded as the wormhole interactions
of the continuum theory. Then the model including these interactions
was expected as a possible theory for $c>1$,
but these interactions could not extend the theory to the region
$c>1$ and a new phase of a special susceptibility has been found for $c\leq 1$.
Many other attempts to resolve this problem
have been tried both in the discretized- and in the
continuum-models, but no one has found an unitary-theory extended to 
the case of $c>1$ without the instability of the tachyon.
Here we approach to this problem from the 
non-critical string theory by taking account of the loop-corrections.
From the viewpoint of the 2d gravity, this approach is equivalent to 
consider the interacting many universes. In this sence, this approach
is similar to the matrix model mentioned above. But a new 
degrees of freedom related to the open-string state appears
in our continuum approach differently from the matrix model, and this new
freedom plays an important role in extending the theory to $c>1$ region.

Our purpose is to show the existence of a phase where the theory 
is stable even in the region $c>1$. This
phase is found by taking account of the loop-correction of
open-strings whose end points couple to the boundary on the world
sheet of the closed-string. The importance of this kind corrections
in considering the vacuum of the string theory
was pointed out previously in \cite{fis}, and the calculational technique
has been developped in ~\cite{fta,clny} in the superstring theory.
Here this technique
is applied to the non-critical string case. In this case, we must take
care about the Liouville-field part in the world-sheet
action.

The analysis is performed around the 
linear dilaton vacuum, but the tachyon might be shifted by a constant
due to the loop-correction.
And we assume that the string coupling constant $g_s$
is small. Namely, we are considering
in the small coupling region.
The guiding principle of our analysis
is the conformal invariance or the BRST invariance of the theory.

\section*{2. Tree Vacuum State}\scs{2}

We set up a vacuum state as a basis to calculate the string-loop
corrections. Such a vacuum
is obtained by imposing the conformal invariance on
the world-sheet action, which is written in the following nonlinear
$\sigma$-model,
\begin{equation}
             S_{\rm 2d}={1 \over 4\pi}\int\,d^2z\sqrt{\hg}
            \left[ {1 \over 2}G_{\mu\nu}(X)\hg^{\alpha\beta}
              \partial_\alpha X^{\mu}\partial_\beta X^{\nu}
           +{\hat R}\Phi(X)+T(X) \right]+\hat{S}_{\rm gh} , 
                           \label{eq:t1} 
\end{equation}
where the ghost action $\hat{S}_{\rm{gh}}$ is expressed as,
$$
           \hat{S}_{\rm gh}={1 \over 2\pi}\int\,d^2z\sqrt{\hg}
              \hat{g}^{\alpha\beta}c^{\gamma}\nabla_{\alpha}
              b_{\beta\gamma}\ \ . 
$$
Here the theory is quantized by the conformal gauge, 
$g_{\alpha\beta}=\rm{e}^{2\phi}\hat{g}_{\alpha\beta}$, where
$\hat{g}_{\alpha\beta}$ is some
fiducial metric. The conformal mode ($\phi$) and $c$-scalar fields
($x^i$, $i=1\sim c$) in the world sheet action are denoted by the target space
coordinates, $X^{\mu}=\left\{\phi, x^i\right\}$, where $\mu=0,i$. 
The naive form of the target-space action is obtained from the 
$\alpha'$-expansion as ~\cite{clty,coop},
\beq
 S_{\rm T}={1 \over 4\pi}\int d^dX\sqrt{G}{\rm e}^{-2\Phi}
          \left\{R-4(\nabla \Phi)^2+(\nabla T)^2-2T^2-{25-c \over 3}
            \right\}, \label{eq:t2}
\eeq
where $d=1+c$. And we do not include the nonlinear
terms coming from $T$-expansion ~\cite{ito,coop} since it gives a minor
change in our analysis and does not affect on the conclusion as discussed
below. While, it might be possible to consider that
\rf{eq:t2} represents the action
obtained after the field redefinitions to leave $T^2$ as the
only possible non-derivative term of $T$ ~\cite{tsey,bank}.
However there is an argument \cite{belo} that it is impossible to rewrite
the power-series of $T$ in such a way. But this point is not
important here.

Solving the equations derived from \rf{eq:t2}, we obtain the following
vacuum solution,
\beq
  G_{\mu\nu}=G_{\mu\nu}^{(0)}\equiv\delta_{\mu\nu}, \,\,
 \Phi=\Phi^{(0)}\equiv {1 \over 2}Q\phi, \,\, T=0, \label{eq:t4}
\eeq
where $Q=\sqrt{(25-c)/3}$. Then,
$S_{\rm 2d}$ can be written as
\begin{equation}
      S_{\rm 2d}={1 \over 8\pi}\int\,d^2z\sqrt{\hg}
            \left[\hg^{\alpha\beta}
              \partial_\alpha X^{\mu}\partial_\beta X_{\mu}
           +Q{\hat R}\phi \right]
           +\hat{S}_{\rm gh}, \label{eq:t5}
\end{equation}
Although \rf{eq:t2} is not an exact form of the target-space action
since higher derivative terms are neglected, but the solution \rf{eq:t4}
is exact in the sence that \rf{eq:t5} is exactly conformal invariant.
In terms of \rf{eq:t5}, the mode expansion of the fields $X^{\mu}$
is performed by noticing that the Coulomb gas picture \cite{dot}
is applicable to the Liouville mode $\phi=X^0$ which couples to 
the background charge. It is obtained as follows,
\bea
 X^{\mu}(z,\bar{z})&=&\varphi^{\mu}(z)+{\bar \varphi}^{\mu}(\bar{z}),
                      \label{eq:28} \\
        \partial \varphi^{\mu}(z)&=&-i\sum_m \alpha_m^{\mu}z^{-m-1},
                      \label{eq:29} \\
   \bar{\partial}{\bar \varphi}^{\mu}(\bar{z})
                   &=&-i\sum_m \bar{\alpha}_m^{\mu}\bar{z}^{-m-1},
                      \label{eq:30}
\eea
where $z=\exp (\tau+i\sigma)$ and
\beq
  [\alpha_m^{\mu}, \alpha_n^{\nu}]=\delta^{\mu\nu}m\delta_{m+n,0}.
                     \label{eq:31}
\eeq
The vacuum for these bosonic fields is defined as,
\beq
 \alpha_m^{\mu}|0>=\bar{\alpha}_m^{\mu}|0>=\left\{\begin{array}{ll}
             0    & \mbox{for $m>0$ and $m=0$, $\mu=i>0$} \\
             -{i \over 2}Q & \mbox{for $m=\mu=0$}\end{array}\right .
                     \label{eq:34}
\eeq
For the ghost fields, we obtain
\bea
        c(z)&=&\sum_m c_m z^{-m+1},
                      \nonumber \\
        b(z)&=&\sum_m b_m z^{-m-2}, \qquad
        \{c_n,b_m\}=\delta_{n+m,0}. \label{eq:38}
\eea
Similar formula are obtained for $\bar{b}(\bar{z})$ and
$\bar{c}(\bar{z})$.

The vacuum of the ghost would be given in the
next section. And the stress tensors for each field
are obtained as follows,
\bea
     T^{\phi}(z)&=&-{1 \over 2}:\partial\phi\partial\phi:
                       +{Q \over 2}\partial^2\phi,  \label{eq:40} \\
     T^X(z)&=&\sum_{i=1}^c -{1 \over 2}:\partial X^i \partial X^i :,
                       \label{eq:41} \\
     T^{bc}(z)&=& :c\partial b+2\partial c b:.
                       \label{eq:42}
\eea
These formula are used to construct the boundary state in the next section.

\section*{3. Boundary State}\scs{3}

Here we construct the loop-amplitude of the open-string.
This is achieved by connecting the tube, which is the 
propagator of the closed string,
to the boundary on the surface of the world sheet since
this tube can be identified as
the loop of the open-string whose end point is sewing 
the boundary of the world surface. Another end point disappears in the vacuum
or couples to the external string configurations.
The boundary state with a tube made in this way
is not confromal invariant and it leads to a conformal anomaly when
the closed-string state propagating through the tube is massless.
We consider this formulation for
the non-critical string case by applying the procedure given for
the superstring case \cite{clny}.

First, consider the boundary (at time $\tau$ and boundary coordinate $\sigma$) 
on the world sheet, where the
following boundary action is assumed,
\beq
   S_B=-{i \over 4\pi}\int_{\partial M}d\sigma 
                 A_{\mu}(X)\partial_{\sigma}X^{\mu} \, . \label{eq:b1}
\eeq
Here $\partial M$ represents the boundary (or boundaries) on the
2d manifold $M$, and
$A_{\mu}(X)$ is the gauge field living on the open-string and
on the boundary. We can consider this gauge field in two ways; (i) It is
the usual gauge field defined on the open string. In this case,
$S_B$
could contain other fields of open-string states, but we concentrate
here our attention on this gauge field only. (ii) It is an auxiliary field
which is needed for the gauge invariance on the boundary
of the antisymmetric field $B_{\mu\nu}$ defined
on the world sheet as, 
$\int d^2z\epsilon^{ab}B_{\mu\nu}\partial_aX^{\mu}\partial_bX^{\nu}$.

In any case,
we assume here that its field strength,
$F_{\mu\nu}=\partial_{\mu}A_{\nu}-\partial_{\nu}A_{\mu}$,
is varing very slowly with $X^{\mu}$.
In this case, $S_B$ can be approximated as
\beq
   S_B={i \over 8\pi}F_{\mu\nu}\int_{\partial M}d\sigma 
                 X^{\mu}(X)\partial_{\sigma}X^{\nu} \, . \label{eq:b2}
\eeq
And the following boundary conditions are needed
\beq
    \left. \partial_{\tau}X^{\mu}+iF_{\mu\nu}\partial_{\sigma}X^{\nu}
          \right |_{\tau}=-{i \over 2}Q \delta_0^{\mu} \, .
                              \label{eq:22}
\eeq
The condition for $\mu=0$ in \rf{eq:22} represents the one
for the Liouville field. It is consistent with \rf{eq:34}, and
it can be corresponded to the boundary
condition for the non-critical open-string ~\cite{Dur,mans}.

The condition \rf{eq:22}
leads to the following operator relations,
\beq
 \alpha_{-m}^{\mu}=-\left({1-F \over 1+F}\right)^{\mu\nu}
\bar{\alpha}_m^{\nu}{\rm e}^{-2m\tau} \label{eq:32}
\eeq
for $\rm{m}\neq 0$ and
\beq
 \alpha_0^i=\bar{\alpha}_0^i=0, \quad
 \alpha_0^0=\bar{\alpha}_0^0=-{i \over 2}Q \,. \label{eq:322}
\eeq
By determining the normalization of the boundary state
according to the path integral method
\cite{clny} with our boundary action \rf{eq:b1},
we obtain the following boundary state for the bosonic part,
\beq
 |B>_{\rm boson}=\sqrt{{\rm det}(G+F)}
\exp (\sum_{m=1}^{\infty}\rm{e}^{2m\tau}\alpha_m^{\mu}
\left({1-F \over 1+F}\right)_{\mu\nu}\bar{\alpha}_m^{\mu})|0>, \label{eq:33}
\eeq
where the bosonic vacuum is defined in \rf{eq:34}.

The operator relations for the ghost part are derived from
the requirement of the BRST invariance of the boundary state,
$(d+\bar{d})|B>=0$. The BRST operator $d$ is defined as 
\bea
     d&=&{1 \over 2\pi i}\oint J(z)dz\,, \nonumber \\
     J(z)&=&:[T^{\phi}(z)+T^X(z)+{1 \over 2}T^{bc}(z)]c(z):\, ,
                \label{eq:39} 
\eea
and $\bar{d}$ can be obtained similarly.
As a result, the following relations are obtained,
\beq
  c_n=-\bar{c}_{-n}\, , \,\,\qquad b_n=\bar{b}_{-n} \, . \label{eq:48}
\eeq
Then the boundary state for the ghost part is obtained as
\beq
 |B>_{\rm gh}=\exp\left\{\sum_{n=1}^{\infty}{\rm e}^{2n\tau}
        [\bar{c}_{-n}b_{-n}+c_{-n}\bar{b}_{-n}]\right\}
         (c_0+\bar{c}_0)|\downarrow\downarrow>, \label{eq:49}
\eeq
where $|\downarrow\downarrow>$ is the Siegel vacuum \cite{clny}
and $<\uparrow\uparrow|\downarrow\downarrow>=1$. 
While the left eigenvector, which is consistent with \rf{eq:48} for $n<0$,
is found as,
\beq
  <\uparrow\uparrow|(b_0-\bar{b}_0)
       \exp\left\{\sum_{n=1}^{\infty}{\rm e}^{2n\tau}
        [\bar{c}_{n}b_{n}+c_{n}\bar{b}_{n}]\right\}
                         , \label{eq:51}
\eeq
In this way, we obtain the boundary state $|B>$ as
$$ |B>=|B>_{\rm boson}|B>_{\rm ghost} $$.

The next step is to add the cylinder, the propagator of the closed-string,
$[L_0+\bar{L}_0-2]^{-1}$,
to the boundary. In order to have a non-zero cylinder-amplitude for the
vacuum boundary state, 
we demand that
the propagator should be accompanied with the zero modes of $b$
in the form, $-(b_0+\bar{b}_0)$ \cite{clny}.
Then, we arrive at the following state with a tube of the closed
string, 
\beq
   |\Psi>_{\rm B}=-(b_0+\bar{b}_0)
  [L_0+\bar{L}_0-2]^{-1}\left\{|B>+|C>\right\}, \label{eq:53}
\eeq
where $|C>$ denotes the boundary state of Mobius strip which gives
a different normalization coefficient
from that of the annuls $|B>$. 

It is easy to derive the relation,
\beq
   (d+\bar{d})|\Psi>_B=|B>_0+|C>_0, \label{eq:54}
\eeq
where $|B>_0$ and $|C>_0$ denote the zero-mass closed-string
state of $|B>$ and $|C>$ respectively. Then if the tachyon is massless, 
we obtain
\beq
   |B>_0+|C>_0=\kappa\sqrt{\rm{det}(G+F)}
     (c_0+\bar{c}_0)|\downarrow\downarrow>, \label{eq:56}
\eeq
where $\kappa$ denotes the product of the
string coupling constant and the numerical
factor depending on the details of the open-string model considered here.

\section*{4. Loop Corrected Phase}\scs{4}

\par

Before examining the loop-corrected action, we consider the tree level action.
The mass-square of the tachyon-field is derived from \rf{eq:t2}
with the vacuum \rf{eq:t4}, and it
is given as follows,
\beq
 m_T^2={1-c \over 12}\, . \label{eq:m1}
\eeq
It becomes negative for $c>1$ as is well-known. 
However, this situation can be changed if the improved
action is considered instead of \rf{eq:t2}, and it is possible to
obtain massless tachyon even if c is larger than one as shown below.
The reason why this is possible is that
the loop-correction given above modifies
the tachyon part of the target-space action if we assume
the masslessness of the tachyon in the considering region of $c(>1)$. 
And the consistency of this
assumption is assured by examining the tachyon mass which is derived
from the corrected target-space action.

 The improved action is obtained so that the loop-corrected
field equations are derived from this action.
The corrected equations of the string-fields are obtained
by demanding the cancellation of the conformal anomalies between the one
coming from the loop-correction and the one obtained
by the usual $\alpha'$-expansion. Then we obtain the following equation
of the BRST invariance,
\beq
   (d+\bar{d})|\Psi>=(d+\bar{d})(|\Psi>_{\rm T}+|\Psi>_{\rm B})=0\ , 
                                    \label{eq:e3}
\eeq
where $|\Psi>_{\rm T}$ represents the fluctuation around the vacuum
\rf{eq:t4},
\beq
   |\Psi>_{\rm T}=\left\{T(x)+h_{\mu\nu}(x)\bar{\alpha}^{\mu}_{-1}
          \alpha^{\nu}_{-1}+\tilde{\Phi}(x)[\bar{c}_{-1}b_{-1}
           +c_{-1}\bar{b}_{-1}]\right\}|\downarrow\downarrow>. \label{eq:e4}
\eeq
It is noticed here that the equations obtained from $\alpha'$-expansion 
for the string fields, $T$, $h_{\mu\nu}$ and $\tilde{\Phi}$, are
derived by the equation, $(d+\bar{d})|\Psi>_{\rm T}=0$.
And it implies the target-space action $S_T$ of \rf{eq:t2}.
While eqs.\rf{eq:e3} and \rf{eq:56} lead to the following equations,
\bea
   (-\bar{\nabla}^2+{1 \over 4}Q^2-2)T&=&-\kappa\rm{det}^{1/2}(1+F), 
\label{eq:e6} \\
   (-\bar{\nabla}^2+{1 \over 4}Q^2)h_{\mu\nu}&=&
         (-\bar{\nabla}^2+{1 \over 4}Q^2)\tilde{\Phi}=0. \label{eq:e7}\\
    \partial^{\mu}h_{\mu\nu}&=&Qh_{0\nu}+\partial_{\nu}\tilde{\Phi} .
                       \label{eq:e8}
\eea
where $\bar{\nabla}^2=\sum_{i=1}^c\partial_i^2+(\partial_0-Q/2)^2$ and
the third equation denotes the gauge fixing condition in this scheme.
Here we notice that the zero-th component of the momentum above
has the following correspondence, $p^0=\alpha^0+iQ/2$.
On the other hand, the differential operator has the correspondence, 
$\partial^0=i\alpha^0$. 

These equations can be obtained from the following target space
action,
\beq
 S^{\rm eff}_{\rm T}=S_{\rm T}
        +2\kappa {1 \over 4\pi} \int d^dX\sqrt{\rm{det}(G+F)}
           \tilde{T}, \label{eq:e9}
\eeq
where $S_{\rm T}$ is given in \rf{eq:t2} and we set as
\beq
    \tilde{T}=\rm{exp}(-2\Phi)T. \label{eq:e10}
\eeq
This form of $\tilde{T}$ is the simplest one, but 
it should be noticed that there are many other possibilities of the form of
$\tilde{T}$ since 
the necessary condition
of $\tilde{T}$ being satisfied is 
$\tilde{T}=\rm{exp}(-2\Phi^{(0)})T$ when we set 
the dilaton-field by the vacuum, $\Phi=\Phi^{(0)}$.
For example, we might take as
$\tilde{T}=\rm{exp}(-\Phi-\Phi^{(0)})T$ \cite{gho}.
Then other conditions would be necessary to remove the ambiguity
of the the form of $\tilde{T}$.
In the case of superstring theory, we should take as 
$\kappa\tilde{T}=\rm{e}^{-\Phi}T$ since $\rm{e}^{\Phi}$ represents
the string coupling constant which is denoted here by $\kappa$. But
we are considering here the vacuum, $\Phi=\Phi^{(0)}$, so the coupling
constant part is separated here.

This modified action includes the gauge field $A_{\mu}$,
so we must obtain the solution of its variational equation,
$\delta S_{T}^{\rm{eff}}/\delta A_{\mu}=0$, simultaneously with 
$G_{\mu\nu}$, $\Phi$ and $T$. If we consider $A_{\mu}$ as the auxiliary
field as mensioned in the previous section, we must integrate it. So the
procedure given here is considered as the integration by a saddle point
approximation.
The equation for $A_{\mu}$ is solved
under the following ansatz,
\beq
  G_{\mu\nu}=G_{\mu\nu}^{(0)}\equiv\delta_{\mu\nu}, \,\,
 \Phi=\Phi^{(0)}\equiv {1 \over 2}Q\phi, \,\, T=\rm{const.}, \label{eq:t44}
\eeq
Under this ansatz, the variational equation is written as
$$ -QF_{0\nu}+\sqrt{\rm{det}(1+F)}\partial_{\mu}
               \left({F_{\mu\nu} \over \sqrt{\rm{det}(1+F)}}\right)=0 \, . $$
Here we search for the solution restricting it to the constant field-strengh,
which is denoted by $\tilde{F}_{\mu\nu}$. Then we obtain
the following solution,
\begin{equation}
  \tilde{F}_{0i}=0,\quad \tilde{F}_{ij}=\rm{const.}
             \label{eq:sol1} 
\end{equation}
and $i,j\neq 0$.
The solution \rf{eq:sol1} is corresponding to the magnetic-field
condensation in the four dimensional case.
Then we substitute \rf{eq:sol1} into $S_{T}^{\rm{eff}}$, 
and the effective potential for $T$ is obtained as,
\beq
     v_{\rm eff}(T)=-2T^2+\tilde{\kappa}T, \label{eq:po1}
\eeq
where $\tilde{\kappa}=2\kappa\sqrt{\rm{det}(1+\tilde{F})}$.

It should be noticed that
this result is the same form with the one obtained without the gauge field if
we consider $\tilde{\kappa}$ as the modified string coupling constant.
Since this potential contains the linear term, $T$ should be shifted by
a constant, $T_1$, to remove the tadpole of the tachyon.
$T_1$ is obtained by solving the equation, $v'_{\rm eff}(T_1)=
dv_{\rm eff}/dT|_{T=T_1}=0$, and we find $T_1=\tilde\kappa/4$. 
And the mass of the tachyon is obtained as
\beq
 m_T^2={25-c \over 12}+{1 \over 2}v^{''}_{\rm{eff}}(T_1)
           ={1-c \over 12}. \label{eq:m21}
\eeq
This is the same with \rf{eq:m1}, so it is impossible to
extend the magnetic-field condensed phase to the region $c>1$.
If we consider other form of potential, 
which is improved at the tree level and 
includes $T^3$ ~\cite{ito,coop} in \rf{eq:t2}, then we find a
mass shift of $T$ of the order $\kappa$ of \cite{gho}.
Then it might be possible to exceed $c=1$ in this phase. In fact,
if we add $T^3/6$ term to \rf{eq:t2} for example, we obtain
$m_T^2={1-c \over 12}+\tilde{\kappa}/2$. Then $c=1+6\kappa$ is
obtained if we set $m_T=0$. However,
it seems to be impossible to go some finite value of $c(>1)$ within
the small-$\kappa$ approximation.
As for the solutions of non-constant field strength, we do not examined
about them here and the problem related to those solutions is remained open.
Here we consider another possibility which is obtained by the condensation
of the dual field, whose definition is given below,
of $A_{\mu}$.

In order to search for a new phase, we rewrite
\rf{eq:e9} in terms of the dual field. It is enough to consider only the
second term of \rf{eq:e9} to do this, and
it can be rewritten by introducing the auxiliary fields,
$\Lambda_{\mu\nu}$ and $f_{\mu\nu}$ as follows,
\[
    \rm{exp} \left(
   -2\kappa {1 \over 4\pi}\int d^dX\sqrt{\rm{det}(G+F)}{\rm e}^{-2\Phi}
           T\right)= \qquad\qquad 
\]
\beq
     \int \rm{D}f_{\mu\nu}\rm{D}\Lambda_{\mu\nu}
           \rm{exp}\left\{
   -2\kappa {1 \over 4\pi}\int d^dX\sqrt{\rm{det}(G+f)}{\rm e}^{-2\Phi}T
           +{1 \over 2}i\Lambda^{\mu\nu}
                   (f_{\mu\nu}-2\partial_{\mu}A_{\nu})
\right\}. \label{eq:e11}
\eeq
And the square root in \rf{eq:e11} can be removed in terms of the formula,
\beq
 \int_{-\infty}^{\infty}\rm{e}^{-{a \over x^2}-bx^2}dx
                =\sqrt{{\pi \over b}}\rm{e}^{-2\sqrt{ab}} \, , \label{eq:fo1}
\eeq
as follows,

$ \qquad \qquad
           \rm{exp}\left(
   -2\kappa {1 \over 4\pi}\int d^dX\sqrt{\rm{det}(G+f)}{\rm e}^{-2\Phi}
           T\right) $
\beq
   =\int\rm{D}v
           \rm{exp}\left\{
             -\int d^dX\left[{1 \over 2v^2}\rm{det}(G+f)
            +{1 \over 2}v^2({\rm e}^{-2\Phi}
                  {\kappa \over 2\pi}T)^2\right]
                   \right\} \, .  \label{eq:e12}
\eeq
In this rewriting, the measure of the path integral gains the
factor $\Pi_i({\rm e}^{-2\Phi_i}T_i)$, where $i$ denotes the 
discretized space-time label. But this factor is reguralized out in the 
dimensional scheme since it gives a volume divergent term in the action,
and we can neglect
this factor.
The determinant part, det$(G_{\mu\nu}+f_{\mu\nu})$, is expressed as
\beq
     \rm{det}(G+f)= \cases{
             \rm{det}G+{1 \over 2}f_{\mu\nu}^2 &
                      \rm{for}\, d=2   \cr
           \rm{det}G+f_{\mu\nu}\Omega^{\mu\nu,\alpha\beta}
                f_{\alpha\beta} & \rm{for}\, d=3 \cr } \label{eq:e13}
\eeq
where $\Omega$ is the 3$\times$3 matrix depending on the metric $G_{\mu\nu}$
and it becomes the unit matrix for $G_{\mu\nu}=\delta_{\mu\nu}$. For $d=4$,
the quartic terms of $f_{\mu\nu}$
appear, so it is difficult to integrate over $f_{\mu\nu}$ exactly.
We discuss this case and the cases of $d>4$ afterward
in the weak field approximation, and 
the explicit form of their determinant are not given here.

As a first example,
we show the explicit calculation for d=2. From eqs.\rf{eq:e11} $\sim$
\rf{eq:e13}, we can integrate $f_{\mu\nu}$ by rewriting
the terms depending on it as,
\beq
  {1 \over 4v^2}f_{\mu\nu}^2+{i \over 2}\Lambda^{\mu\nu}f_{\mu\nu}
  ={1 \over 4v^2}(f_{\mu\nu}+iv^2\Lambda_{\mu\nu})^2+
  {v^2 \over 4}\Lambda_{\mu\nu}^2\, . \nonumber
\eeq
After that, we perform the $v$-integration, and we obtain
\beq
   -\int d^dX\left[\sqrt{\rm{det}G}
       \sqrt{({\rm e}^{-2\Phi}{\kappa \over 2\pi}T)^2
                +{1 \over 2}\Lambda_{\mu\nu}^2}
             -i\Lambda^{\mu\nu}\partial_{\mu}A_{\nu}
                   \right] \, .  \label{eq:e15}
\eeq
Then $\Lambda_{\mu\nu}$ is solved by the constraint, 
$\partial_{\mu}\Lambda^{\mu\nu}=0$, which is
obtained from the integration of $A_{\mu}$. The solution is found as,
$$ \Lambda^{\mu\nu}=\epsilon^{\mu\nu}\bar{a}$$
where $\bar{a}$ is an arbitrary constant. Then we obtain the following
correction term,
\beq
   -{2\kappa \over 4\pi}\int d^dX \rm{e}^{-2\Phi}\sqrt{\rm{det}G}
       \sqrt{T^2+({2\pi \over \kappa}\bar{a})^2\rm{e}^{4\Phi}}
              \, .  \label{eq:e16}
\eeq
And the effective action is obtained by replacing the second term of 
\rf{eq:e9} by \rf{eq:e16}. For $d=2$, the dual field is
a constant $\bar{a}$, and we obtain
$\bar{a}=0$ by solving its variational equation.
Then the effective potential of $T$ is obtained as
$$
     v_{\rm eff}(T)=-2T^2+\kappa T \, 
$$
which is the same form with \rf{eq:po1} obtained for $F_{\mu\nu}$ 
condensation phase. Then the tachyon is massless and
the consistency with the calculation given here
is assured. But we can not find any difference from the magnetic field
condensation given above.

Since our interest is in the region of $c>1$, then we consider the three
dimensional case ($c=2$) nextly. After the procedure similar to the 2d case,
we obtain the following correction term for $d=3$,
\beq
   -2\kappa {1 \over 4\pi}\int d^dX \rm{e}^{-2\Phi}\sqrt{\rm{det}G}
       \sqrt{T^2+({\pi \over \kappa})^2
              \Lambda_{\mu\nu}^0\Omega^{\mu\nu,\alpha\beta}
              \Lambda_{\alpha\beta}^0\rm{e}^{4\Phi}
}\, , \label{eq:e17}
\eeq
where 
\beq
      \Lambda^{\mu\nu}_0=\epsilon^{\mu\nu\lambda}\partial_\lambda a(X)\, ,
                                \label{eq:e18}
\eeq
and $a(X)$ is the dual field, which is not a constant but a scalar function
in this case.
This scalar field $a(X)$ is solved by the following variational equation,
\beq
   \partial_k\left({\rm{e}^{2\Phi^{(0)}}\partial_k\it{a} \over
       \sqrt{T^2+({2\pi \over \kappa})^2
              (\partial_{\mu}a)^2\rm{e}^{4\Phi^{(0)}}}}\right)=0
\, . \label{eq:sa1}
\eeq
This equation is obtained by substituting $G_{\mu\nu}=\delta_{\mu\nu}$,
$\Phi=\Phi^{(0)}$ into the original variational-equation since we are
restricting the type of the solution to this kind. Further, we assume
the constancy of $T$ and the condensation of the dual field in the form,
\beq
   \tilde{a}_k^2=\tilde{a}_k\tilde{a}^k=({\Lambda_0 \over 2\pi})^2
           \, , \label{eq:c21}
\eeq
where $\tilde{a}_k=\rm{e}^{2\Phi^{(0)}}\partial_k\it{a}$.
Then \rf{eq:sa1} is written as,
\beq
   Q\partial_0a+\partial^2a=0 \, . \label{eq:c20}
\eeq
A simple solution of this equation is obtained as follows,
\beq
     a=\rm{e}^{\beta\phi}f(X_i), \qquad \partial_i^2f(X_i)
               =-\beta(\beta+Q)f \, ,
                    \label{eq:c19}
\eeq
where $\beta$ is a constant and $f(X_i)$ is a function of $X_i$ 
with $i\neq 0$. Although it is easy to solve $f(X_i)$ and
to write its general solution, but we abbreviate it since
it is not necessary hereafter. The important is the fact that
there is an explicit solution of \rf{eq:sa1} with the condition
of \rf{eq:c21}. Taking into account of \rf{eq:c21},
we arrive at the following effective potential,
\beq
     v_{\rm eff}(T)=-2T^2+2\kappa\sqrt{T^2+({\Lambda_0 \over \kappa})^2} \, .
                     \label{eq:po2}
\eeq

According to the discussion given above we must find the shift ($T_1$)
of the tachyon from the above potential.
It is given by $v'_{\rm eff}(T_1)=dv_{\rm eff}/dT|_{T=T_1}=0$. 
We find two branches of the solution, 
\beq
     T_1= \cases{
             0 & \rm{(A)}\,    \cr
        {\kappa\over 2}\sqrt{1-\lambda^2} & \rm{(B)}
            \cr}  \label{eq:br2}
\eeq
where $\lambda\equiv 2\Lambda_0/\kappa^2$. Here we assume that
$\Lambda_0$ is the order of $\kappa^2$ so that $\lambda$ becomes finite.
As for $T_1$, it is the order of $\kappa$ or zero. 
This setting is consistent with our approximation of the calculation.
For these solutions, the tachyon-mass is obtained as
\beq
 m_T^2={25-c \over 12}+{1 \over 2}v_{\rm{eff}}^{''}(T_1) \nonumber 
\eeq
\beq
     = \cases{
       {1-c \over 12}+{2 \over \lambda} & \rm{(A)}\,    \cr
       {1-c \over 12}+2\lambda^2 & \rm{(B)}
            \cr}  \label{eq:m2}
\eeq
Since the above caluculation has been done for $d=3$ ($c=2$) and $m_T=0$,
then we obtain
\beq
     \lambda= \cases{
             {24 \over c-1}=24 & \rm{(A)}\,    \cr
        \sqrt{{c-1\over 24}}={1 \over 2\sqrt{6}} & \rm{(B)}
            \cr}  \label{eq:br3}
\eeq
for each branch. Both cases are acceptable as far as we consider the
$d=3$ case. While, if we try to extend the above results to the
$c=1$ limit, the branch $(B)$ is smoothly connected to the previous
$c=1$ result, $T_1=\kappa/2$ and $\Lambda_0=0$ at the limit $\lambda=0$.
On the other hand, the branch $(A)$ produces a large gap at c=1 with the
solution obtained at $d=2$.
In this sence, the $(B)$-branch is preferable in order to extend
this phase to the wider region of $c$, and it
seems to be expandable to the region
of $25\ge c\ge 1$. The upper bound of $c$ comes from the reality 
condition of $T_1$.
We can interpret this phenomenon
as a phase transition where the transition point is
$c=1$ and the order parameter is not $T_1$ but
$\Lambda_0$. 

Of course, there are other solutions of \rf{eq:sa1} which do not lead
to the potential \rf{eq:po2}. One simple example is obtained as follows.
Rewrite \rf{eq:sa1} as
\beq
  \partial_kE^k=0 \, , \label{eq:sa22}
\eeq
where $E_k=\tilde{a}_k/ \sqrt{T^2+({2\pi \over \kappa})^2\tilde{a}_k^2}$
and $\tilde{a}_k=\rm{e}^{2\Phi^{(0)}}\partial_k\it{a}$ which is given
above. Then the solution is obtained as
$$ E^k=\epsilon^{kij}\partial_i\eta_j$$
where $\eta_i$ is an arbitrary function, and we obtain the following
equation,
\beq
  \tilde{a}_k^2=T^2{E_k^2 \over 1-({2\pi \over \kappa})^2E_k^2} \, .
                \label{eq:c22}
\eeq
Here, 
\bea
  E_k^2&=&E_kE^k={1 \over 2}f_{ij}^2(\eta) \nonumber \\
    f_{ij}(\eta)&=&\partial_i\eta_j-\partial_j\eta_i \, . \label{eq:c23}
\eea
Next, we assume the condensation of $f_{ij}(\eta)$ such that
$$ <E_kE^k>=\alpha^2$$
where $\alpha$ is a constant. Then we obtain,
\beq
  \tilde{a}_k^2=T^2{\alpha^2 \over 1-({2\pi \over \kappa})^2\alpha^2} \, .
                \label{eq:c24}
\eeq
Substituting this into the target-space action, we get the same form of
potential with \rf{eq:po1} but the coefficient of 
the linear term of $T$ is different.
So this solution leads to the negative mass-squared of the tachyon.

Now we turn to the case of $d=4$ (or $c=3$), we obtain in this case
\beq
     \rm{det}(G+f)= \rm{det}G+f_{\mu\nu}\omega^{\mu\nu,\alpha\beta}
                f_{\alpha\beta}+f_{\rm{qrt}}\, ,\label{eq:e20}
\eeq
where $\omega$ is 6$\times$6  matrix written by the metric $G_{\mu\nu}$,
and $f_{\rm{qrt}}$ represents the quartic terms of $f_{\mu\nu}$. 
Eq.\rf{eq:e20} can be explicitly
written for the flat metric, $G_{\mu\nu}=\delta_{\mu\nu}$, as follows
\beq
     \rm{det}(G+f)= \rm{det}G+{1 \over 2}f_{\mu\nu}^2
                +\left({1 \over 2}\epsilon_{ijk}
                f^{0i}f^{jk}\right)^2\, ,\label{eq:e201}
\eeq
where $i,j,k=1\sim 3$. Due to the quartic term of $f_{\mu\nu}$,
it is difficult to perform the 
complete integration over $f_{\mu\nu}$. 
Then we concentrate our attention on the weak field region
where $f_{\mu\nu}$ is small and the quartic term
can be neglected compared to
the quadratic one. In this case, we can proceed the similar calculation 
to the one performed in
$d=2,3$ cases, and the same form of the result with \rf{eq:e17} is obtained
by replacing the matrix $\Omega$ and $\Lambda^{\mu\nu}_0$ in \rf{eq:e17}
by $\omega$ and
\beq
      \Lambda^{\mu\nu}_0=\epsilon^{\mu\nu\lambda\sigma}
              \partial_\lambda \tilde{A}_{\sigma}\, ,
                                \label{eq:e181}
\eeq
respectively. Here $\tilde{A}_{\mu}$ represents the dual gauge field.
Similarly to the case of $d=3$, we obtain the following variational
equation with respect to $\tilde{A}_{\mu}$,
\beq
   \partial_{\mu}\left({\rm{e}^{2\Phi^{(0)}}\tilde{f}_{\mu\nu}\over
       \sqrt{T^2-{2\pi \over \kappa}^2{1 \over 8}
              (\tilde{f}_{\mu\nu})^2\rm{e}^{4\Phi^{(0)}}}}\right)=0
\, , \label{eq:sb1}
\eeq
where $\tilde{f}_{\mu\nu}=
\partial_{\mu}\tilde{A}_{\nu}
-\partial_{\nu}\tilde{A}_{\mu}$.
As in the previous case, we solve this equation under the condition
of constant $T$ and
\beq
   \rm{e}^{2\Phi^{(0)}}\tilde{f}_{\mu\nu}=\epsilon_{\mu\nu\lambda\sigma}
         \bar{f}^{\lambda\sigma}
       \, , \qquad
          \pi^2\bar{f}_{\mu\nu}^2=\Lambda_0^2 \, ,
                                                    \label{eq:sb2}
\eeq
where $\Lambda_0$ is a constant. This ansatz is taken so that we can
obtain the same form of potential with \rf{eq:po2}. 
Then \rf{eq:sb1} is written as
$$ \epsilon_{\mu\nu\lambda\sigma}\partial_{\mu}\bar{f}^{\lambda\sigma}
   =0$$
Then a simple solution is obtained as 
$\bar{f}_{\mu\nu}=
\partial_{\mu}\bar{A}_{\nu}
-\partial_{\nu}\bar{A}_{\mu}$,
where $\bar{A}_{\mu}$ is an arbitrary function.
Since we do not need the explicit
form of $\bar{A}_{\mu}$, it is not necessary to specify it here
as in the d=3 case. The important is the fact that
we can arrive at the same effective-potential with \rf{eq:po2}, and
the values of $T_1$ and $m_T$ are given by the same formula with 
\rf{eq:br2} and \rf{eq:m2}. The difference is the value of
$c$, which is taken at $c=3$ here. As in the case of $c=2$,
both branches are possible, but
we prefer (B)-branch since it can be connected to $c=1$ as stated above. 
As a result,
the massless tachyon phase is realized also in $d=4$ due to the 
condensation of the dual field.

Within the weak field approximation, it is straightforward to extend the
analysis of $d=4$ to the cases of $d>4$ by taking only the quadratic term
of $f_{\mu\nu}$ into account.
In this way, we can extend the formula \rf{eq:po2} $\sim$ 
\rf{eq:br3} to the region of
$d\ge 4$ or $c\ge 3$ within the weak field approximation.

Finally we remind the vacuum configuration of the other fields. 
It is derived from $S^{\rm eff}_{\rm T}$ as follows, 
\beq
 G_{\mu\nu}=\delta_{\mu\nu}, \,\,
 \Phi={1 \over 2}Q\phi, \,\, T=T_1 \ . \label{eq:g15}
\eeq
Here $Q$ and $T_1$ vary with $c$, but this form of \rf{eq:g15}
is not changed for $c\geq 1$.
We should notice that non-zero value of $T_1$ is not essencial to the vacuum 
structure. The essential point is the value of $\Lambda_0$. 
The origin of $\Lambda_0$ is
however hidden in the target-space action, which controles the dynamics of
the closed string and determines the form of its world-sheet action.
From the viewpoint of the 2d gravity, the above analysis implies the 
importance of the interaction of open universe and 
the closed universe by taking a picture of interacting many universes.
This picture could open the way to arrive a 2d gravitational theory of
$c>1$.

\section*{5. Concluding remarks}\scs{5}

Quantum fluctuation, which is corresponding
to the loop-correction of the open string, of 2d surface
is considered to find
a vacuum of the non-critical string theory
for $c>1$. The calculational technique developed in the 
critical super-string theory is applied. The loop-correction is given
in terms of the boundary state accompaning a tube of the closed
string propagator. The important point is that
this correction provides a conformal anomaly for
the massless-state channel of the
closed string states propagating the tube. 
As a result, the field equation of the tachyon (the ground state of the
closed-string)
is modified if the tachyon is assumed to be massless even for
$c>1$. This assumption is justified by obtaining the zero-mass tachyon state
from the effective target-space action which is improved by the
loop-correction.

The essential factor to get this justification is the 
the condensation of the dual field of the gauge field, which is
in the corrected action of the Born-Infeld type.
The gauge field lives on the boundary of the
world-sheet and is also needed from the gauge invariance on the world-sheet
with boundaries for the antisymmetric tensor of the closed-string states.
However, we could not obtain a phase, where there is no c=1 wall,
by the condensation of this gauge field directly. Such a phase
has been found in the vacuum, where the dual field of this gauge field
condenses. 
It is found that
the ground state of the closed string is the massless tachyon in this phase.
This is seen in terms of the dual transformed
target-space action.
For $d=2,3$, the dual transformation is exact.
But, it is performed for $d\ge 4$
by the weak field approximation. Within this approximation,
the phase found here seems continuing to exist
up to the critical dimension $c=25$, where the Liouville field disappears.
So the condensation of the
dual field is essential to this phase where
there is no $c=1$ wall.

\noindent{\bf Acknowledgment:}
The author thanks for the encouragement of T.Yukawa and 
H.Kawai at the conference held in October, 1996 at Shikanosima. He also thanks
K.Inoue for useful discussions.

\newpage

\end{document}